\documentclass[12pt,preprint]{aastex}

\begin{document}
\title{A Statistical Description of AGN Jet Evolution from the VLBA Imaging and Polarimetry Survey (VIPS)}
\author{J. F. Helmboldt\altaffilmark{1}, G. B. Taylor\altaffilmark{2}, R. C. Walker\altaffilmark{3}, and R. D. Blandford\altaffilmark{4}}

\altaffiltext{1}{Naval Research Laboratory, Code 7213, 4555 Overlook Avenue SW, Washington, DC 20375-5351}
\altaffiltext{2}{Department of Physics and Astronomy, University of New Mexico, 800 Yale Blvd NE, Albuquerque, NM 87131, USA}
\altaffiltext{3}{National Radio Astronomy Observatory, P.O. Box O, Socorro, NM 87801, U.S.A.}
\altaffiltext{4}{KIPAC, Stanford University, PO Box 20450, MS 29, Stanford, CA 94309, USA}
\received{?}
\accepted{?}

\begin{abstract}
A detailed analysis of the evolution of the properties of core-jet systems within the VLBA Imaging and Polarimetry Survey (VIPS) is presented.  
We find a power-law relationship between jet intensity and width that suggests for the typical jet, little if any energy is lost as it moves away from its core.  Using VLA images at 1.5 GHz, we have found evidence that parsec-scale jets tend to be aligned with the the direction of emission on kiloparsec scales.  We also found that this alignment improves as the jets move farther from their cores on projected scales as small as $\sim$50-100 pc.  This suggests that realignment of jets on these projected scales is relatively common.  We typically find a modest amount of bending (a change in jet position angle of $\sim 5^{\circ}$) on these scales, suggesting that this realignment may typically occur relatively gradually.

\end{abstract}


\section{Introduction}
While the properties of active galactic nuclei (AGN) have been studied across the entire electromagnetic spectrum, Very Long Baseline Interferometry (VLBI) in the radio regime has provided the only means to study the parsec-scale structures associated with the many AGN.  Chief among these structures are the relativistic jets of material launched by most AGN.  One of the keys to understanding these phenomena is the development of a full picture of their evolution from the smallest ($\sim$10 AU) to the largest (up to $\sim$1 Mpc) scales.  Observations of AGN in the X-ray regime with the Chandra satellite as well as radio images made with interferometers such as the Very Large Array (VLA) have helped define the typical properties of jets and associated lobe emission on large ($\sim$1 kpc) scales.  The High Energy Stereoscopic System (HESS) has provided insights into the gamma-ray emitting regions ($\sim$0.01-1 pc), but these observations are biassed toward one class of AGN, BL Lac objects \citep{aha05}.  Future observations with the Gamma-Ray Large Area Space Telescope \citep[GLAST; ][]{geh99} will provide similar insights, but for a much broader and larger sample of sources.  However, data from both HESS and GLAST provide virtually no spatial information about jets on any size scale, and require radio-regime VLBI images for proper interpretation.  In addition to this, it has been seen that the strongest gamma-ray AGN may be biassed toward radio sources with higher brightness temperatures and larger core fractions \citep{tay07}.  From this, it is clear that radio-frequency VLBI images of AGN are vital to the study of the evolution of the structure and properties of jets on parsec scales.\par
While many VLBI monitoring campaigns have been used to examine jets evolving over time on these size scales \citep[see][and references therein]{lis05}, they have typically been focused on modest samples of the brightest compact sources since monitoring of a relatively large sample of sources may be impractical.  This has precluded any significant statistical analysis of tendencies among a variety of core-jet systems regarding their evolution.  However, it is plausible that single epoch data for a relatively large number of sources can be used to statistically discern how the typical jet evolves as each jet within the sample is presumably in a different stage of evolution.  If a large sample with well defined selection criteria is used, the affects of potential biasses on this analysis can be properly constrained, and a reasonably accurate representation of the typical core-jet system can be constructed.  We have used the 5 GHz images obtained by the VLBA Imaging and Polarimetry Survey \citep[VIPS; ][]{hel07} for 1,127 AGN with the NRAO Very Long Baseline Array (VLBA) to demonstrate that this is possible.  In \S 2, we provide a brief description of the VIPS sample.  In \S 3, we outline the selection biasses that affect our proposed analysis and describe how they have been over-come or, at least, minimized.  In addition, we describe in \S 3 the statistical constraints we have place on the evolution of the intensity, width, and direction of the typical jet using our VIPS data.  In \S 4, we summarize the implications of these results for the evolution of the intrinsic properties of the typical jet.

\section{The VIPS Core-jet Sample}
The VIPS source list was selected from the Cosmic Lens All-Sky Survey \citep[CLASS; ][]{mey03}, an 8.5 GHz VLA survey of $\sim$12,100 flat spectrum objects ($\alpha > -0.5$ between 4.85 GHz and a lower frequency).  VIPS sources were selected to be all CLASS sources brighter than 85 mJy at 8.5 GHz which are also within the survey area of the Sloan Digital Sky Survey \citep[SDSS; ][]{yor00} and with declinations between 15$^{\circ}$ and 65$^{\circ}$.  For 147 of these sources, data were already available at 5 GHz from the Caltech-Jordell Bank Flat spectrum survey \citep[CJF; ][]{tay96,bri03,pol03}; 20 were observed for the VIPS pilot program at 5 and 15 GHz \citep{tay05}.  There are 8 sources not observed as part of either CJF or the pilot program which have been observed at 15 GHz as part of the Monitoring of Jets in AGN with VLBA Experiments project \citep[MOJAVE; ][]{lis05}.  The remaining 958 were observed with the VLBA at 5 GHz in full polarization from January to August 2006.  The data were processed automatically using a combination of the VLBA calibration pipeline \citep{sjo05} and DIFMAP imaging scripts described in detail in \citet{tay05}.  Data for the 147 sources taken from the CJF survey were reprocessed using the same imaging scripts.  All but $<$1\% of the sources were significantly detected; about 36\% had detected polarized intensity.\par
Each source was broken up into components by fitting Gaussian functions to significant ($>6\sigma$) peaks in its 5 GHz image.  These components were used, along with visual inspection, to classify each source as a point-like source, a core-jet system, a compact symmetric object (CSO) candidate, or a complex source.  
In the analysis presented here, only core-jet systems with more than one Gaussian component are considered, i.e., sources where the core and jet have been resolved into separate components, 602 sources in all.\par
In addition to the 5 GHz VLBA data, we also have optical spectra for 559 ($\sim$50\%) of the VIPS sources.  Among the 602 VIPS core-jet systems presented here, 311 have optical spectra.  These were taken from a combination of SDSS spectra and ongoing follow-up observations being conducted with a variety of telescopes \citep[see ][]{hea08}.  The spectra have been used to derive redshifts for these sources which have been used to compute distances assuming ($\Omega_{m}$,$\Omega_{\Lambda}$,$h$)=(0.3,0.7,0.7).  Throughout this paper, all jet lengths and projected separations of jet components from their cores that are presented in physical units (i.e., pc) were computed using this assumed cosmology.

\section{The Evolution of Jet Properties}
While the VIPS data set provides information regarding each source at only one epoch, the relatively large number of sources allows us to explore statistically how the properties of jet components change as they move away from their cores.  To do this, we have computed the projected separation of each jet component from its core as well as its apparent position angle with respect to the core position for all core-jets with known redshifts.  We have also computed the mean brightness temperature for each of these components using the parameters of the Gaussian components fit to the 5 GHz VLBA images.  We note that this allows for a statistical description of how the properties of the typical jet change with projected core separation, but not necessarily with time.  This is because while there is evidence that the jet speed, $\beta$, is relatively constant with core separation for individual jets \citep[see, e.g., ][]{kel07}, the mean value of $\beta$ among our sample of core-jet systems may change.  This caveat will be discussed further in \S 4.

\subsection{Brightness Temperature and Luminosity}
\subsubsection{Brightness Temperature and Luminosity}
To examine how brightness temperature and luminosity evolve on average, we can plot the brightness temperature and luminosity of each jet component as a function of its projected separation from the core in physical units rather than using apparent separations.  Perhaps the most desirable way to do this would be to measure the separations in terms of physically meaningful scale-lengths such as the Schwarzschild radius or the core radius of the host galaxy.  However, we do not have reliable black hole mass estimates for many of our sources \citep[see][]{hel07}, and for many of our higher redshift sources, the host galaxies are unresolved in their optical images.  These facts make it difficult to use such scale-lengths for a useful number of sources.\par
A more straightforward approach is to use the projected separation in units of parsecs.  While this does not take into account the distributions of black hole masses or host galaxy sizes, it provides us with a reasonable means for statistically constraining the evolution of jet components for the typical core-jet system using all of our sources with known redshifts.  When examining how the component brightness temperature and luminosity vary with projected core separation, however, we must deal with sources of bias brought on by the selection criteria for VIPS sources, the sensitivity and angular resolution of the VIPS 5 GHz observations, the sources' orientations, and the effects of cosmology.  These effects are problematic for two reasons.  First, the finite angular resolution of the 5 GHz images ensures that the jet components with the smallest projected separations (in physical units) will be within the most nearby sources.  Second, the components with the largest separations will likely be within the most distant sources.  This is because relatively faint jet emission that lies beyond the  limits placed on the image field of view by time averaging and bandwidth smearing of our VLBA data will not significantly effect the fit to the visibility data obtained using the clean algorithm with DIFMAP, and it will not be included in the final image of the source.  For these two reasons, we must contend with the following five dominant sources of bias:
\begin{enumerate}
\item The VIPS 8.5 GHz integrated flux density limit of 85 mJy will cause jet components within more distant sources, which likely dominate those with the largest projected separations, to be biassed toward more luminous sources which likely have brighter jet components on average.
\item Cosmological effects reduce the intrinsic brightness temperature of each jet component by a factor of $(1+z)$ \citep[e.g.,][]{kel98} which, when combined with the noise level in each 5 GHz image, will bias jet components within more distant sources to higher intrinsic brightness temperatures.
\item Since the typical spectral index (for $F_{\nu} \propto \nu^{\alpha}$) for jet components is roughly $-0.8$, the observed brightness temperatures will be even lower for jet components within sources at higher redshifts.  Combined with cosmological effects, this implies that the brightness temperature of a typical jet component is reduced by a factor of about $(1+z)^{1.8}$.
\item To be able to measure the intrinsic brightness temperature of a jet component, it must be resolved, implying that for redshifts at or approaching 1.605 where the angular diameter distance peaks for our assumed cosmology, there will be a bias toward physically larger components.
\item Jet components within sources with smaller viewing angles will, on average, have larger Doppler factors, $\delta$, and therefore, larger observed brightness temperatures and luminosity densities and possibly smaller than average projected separations from their cores.  This bias could be made stronger if the typical jet speed is larger closer to the core.  However, most observational evidence seems to indicate that jet speeds remain relatively constant \citep[see ][and references therein]{kel07}.
\end{enumerate}
We may construct a sample of jet components that is free of the first four biases listed above by setting a maximum redshift, $z_{max}$, and then selecting those components which (1) have source 8.5 GHz luminosity densities such that they would have $F_{8.5}>85$ mJy at $z=z_{max}$, (2) have observed brightness temperatures $>5\times (3.39\times 10^{8} \overline{\sigma})(1+z_{maz})^{1.8}$ where $\overline{\sigma}$ is the typical rms of the 5 GHz images of 0.24 mJy beam$^{-1}$, and (3) have intrinsic sizes large enough for them to be resolved if they were at a redshift of 1.605 or $z_{max}$, whichever is smaller, all assuming the component viewing angles are not changed.  We have developed such a sample using $z_{max}=2.3$ which yields a limiting intrinsic brightness temperature of $9.83 \times 10^{7}$ K and a limiting 8.5 GHz luminosity density of $3.46 \times 10^{34}$ ergs s$^{-1}$ Hz$^{-1}$.  Among the components from sources with redshifts that meet these criteria, 38 have large enough intrinsic sizes to be resolved at $z=1.605$.\par
Unfortunately, we do not have any data that would allow us to estimate the viewing angles of these 38 components, and therefore cannot eliminate the fifth bias in the list above.  However, since the component brightness temperatures are boosted by a factor of $\delta$ and the luminosity densities are boosted by a larger factor of $\delta^{n}$, where $n \approx $2.8-3.8 (for $\alpha \approx -0.8$), we may use plots of both these quantities versus projected core separation for these components to explore this effect.  We have plotted the intrinsic brightness temperature (i.e., corrected for the effects of cosmology, angular resolution, and spectral steepness) and the rest-frame 5 GHz luminosity density, assuming $\alpha=-0.8$, for these 38 components as functions of projected core separation in the upper and lower panels of Fig.\ \ref{jetcomps}, respectively.  Points representing components from the same sources are connected with dotted lines.  These lines show all four sources with multiple components represented in the plots have brightness temperatures that decrease with increasing projected separation.  Conversely, the luminosity densities of the same components show no tendency for either a positive or negative gradient.  Mean values which are plotted within bins of projected separation also show that, for all components, the typical brightness temperature is about twice as large at the smallest projected separations ($\sim$30 pc) than it is at the larger separations ($^{>}_{\sim}$150 pc), while the typical luminosity density remains roughly constant.  From these results we must conclude that either (1) the effect of the Doppler boosting bias is minimal and the trend seen in the upper panel of Fig.\ \ref{jetcomps} reflects a real trend for the intensity of jet components to decrease as they move farther from their cores, or (2) the trend is mostly caused by the typical Doppler factor being higher among components with smaller projected separations and the intrinsic luminosities substantially increase with increasing core separation.  While our data cannot rule out either scenario or even some combination of the two, we note that both are consistent with a real trend for the physical sizes of the components to expand as they move outward.  In fact, we find a significant correlation between intrinsic jet component size and projected core separation for the 38 components plotted in Fig.\ \ref{jetcomps} with a Spearman rank correlation coefficient of 0.47.
\subsubsection{Brightness Temperature and Jet Width}
\begin{figure}
\plotone{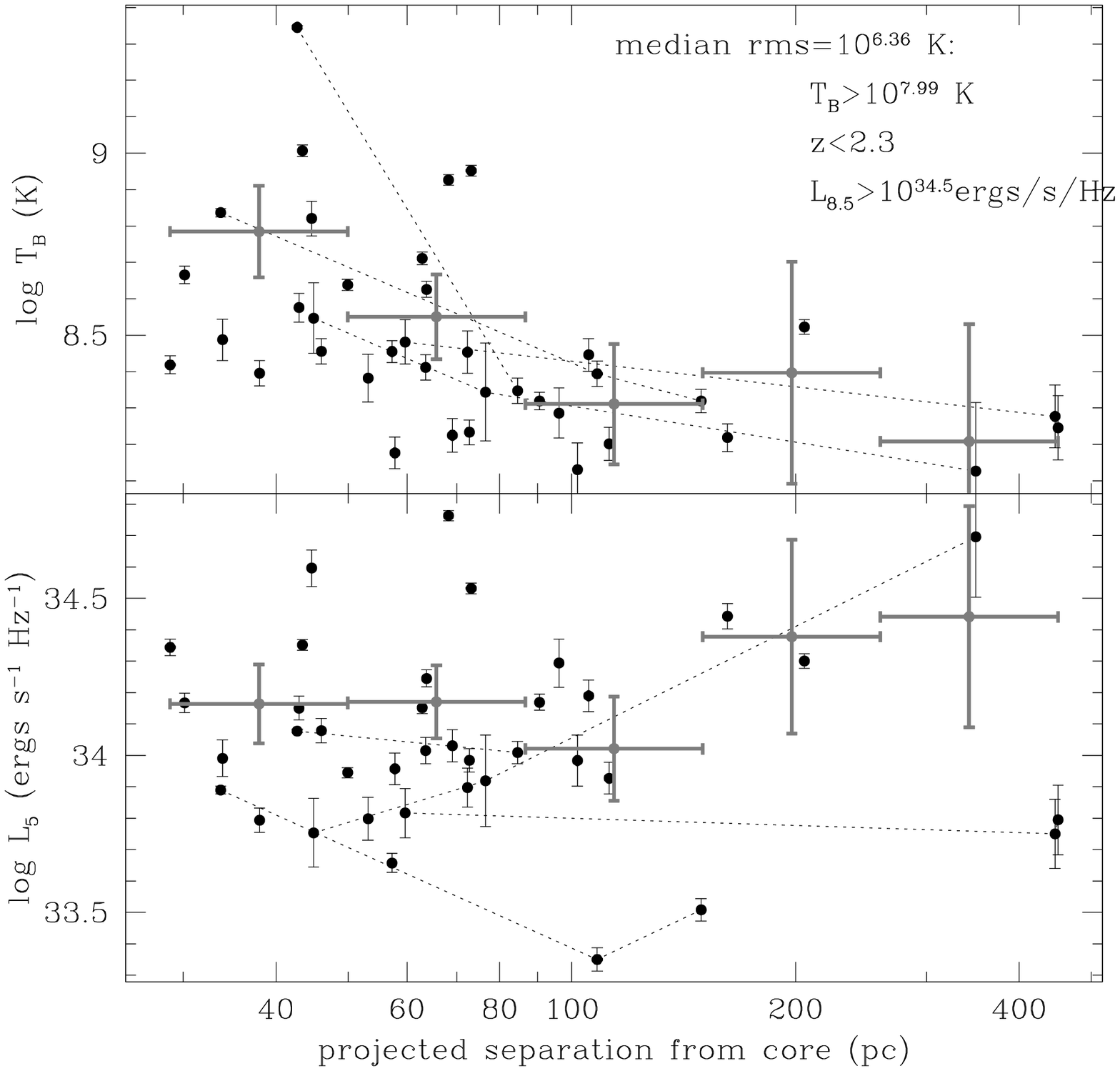}
\caption{The intrinsic brightness temperature (upper) and rest-frame 5 GHz luminosity density (lower) versus projected separation from the core for jet components with intrinsic intensities $T_{B}>10^{7.99}$ and which would be resolved at $z=1.605$ where the angular diameter distance peaks (see \S 3.1) and which have parent sources with 8.5 GHz luminosity densities $>10^{34.5}$ ergs s$^{-1}$ Hz$^{-1}$ and $z<2.3$.  In both panels components belonging to the same source are connected with a dotted line and mean values are plotted within bins of projected core separation as grey points with errorbars.}
\label{jetcomps}
\end{figure}

The results presented in Fig.\ \ref{jetcomps} suggest that individual jet components tend to expand as they move outward, which is by no means a new result.  However, how the intensity of each jet varies with the dimension(s) of the jet can provide insights into the changes in internal energy that may occur as the jet moves outward.  To do this properly, we must consider not just the properties of individual components, but the intensity and width of each jet as a function of core separation.  This is made difficult, however, by the limited sensitivity of our observations which is adequate for identifying the brighter features within each jet, but not for detecting the underly, relatively continuous emission.  To overcome this limitation, we have created a combined mean image, or "stacked" image, of the 31 sources with jet components used in Fig.\ \ref{jetcomps}.  We have limited ourselves to these 31 sources because (1) they belong to a quasi-volume limited sample (i.e., ignoring viewing angle effects), (2) they each have at least one jet component that is detectable on the image out to the chosen maximum redshift of 2.3, and (3) the results presented in Fig.\ \ref{jetcomps} imply that the potential Doppler boosting bias is likely relatively weak for this sub-sample, at least for projected separations $\gtrsim 30$ pc.\par
\begin{figure}
\plotone{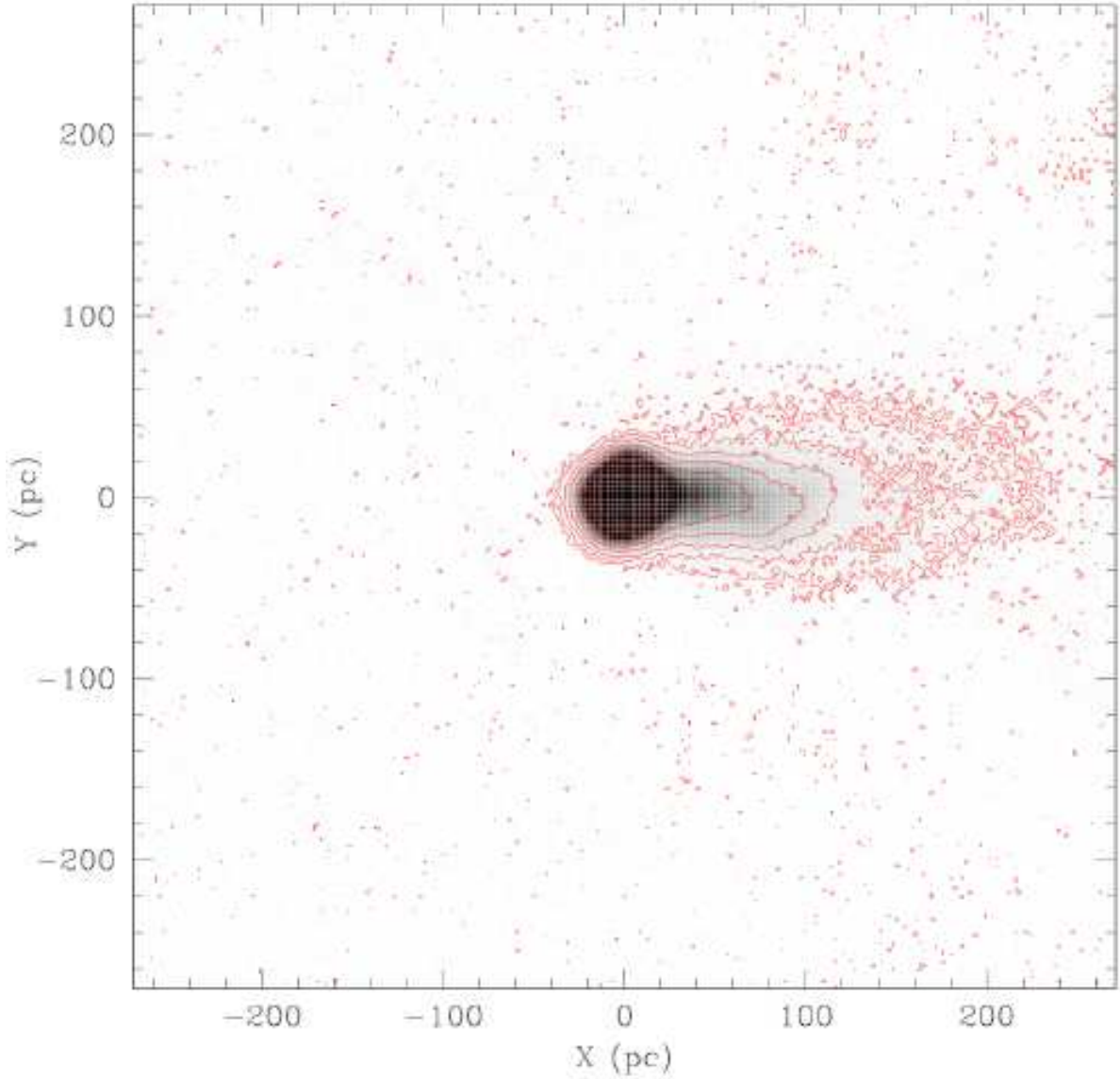}
\caption{The mean rest-frame 5 GHz image for the 31 sources with components plotted in Fig.\ \ref{jetcomps} (see \S 4.1.2) with contours (in red) for intensities of 0.6$\times$(-1, 1, 2, 4, 8, 16, 32, 64, 128, 256, 512, 1024) mJy mas$^{-2}$.}
\label{coaddimg}
\end{figure}

We combined the images by first remaking each image with a pixel width corresponding to 2.12 pc and a restoring beam of $25.4 \times 15.3$ pc, the physical sizes of the nominal pixel size and restoring beam used for a redshift of 1.605.  The phase center for each image was also adjusted to be the position of the source's core.  These new images were then rotated so that the mean jet axis of each was oriented along the x-axis of the image after which each was multiplied by $(1+z)^{1.8}$ and divided by the angular area of its restoring beam to put it in units mJy mas$^{-2}$ at rest-frame 5 GHz.  Following this, an appropriate amount of Gaussian noise was added to each image so that its rms would be nearly equal to the image with the largest rms (1.4 mJy mas$^{-2}$) to ensure the images had roughly the same sensitivity.  The images were then averaged together and the end result can be seen in Fig.\ \ref{coaddimg}.  It can be seen that the jet in this image shows no clear signs of distinct components and predominantly consists of continuous emission.\par
\begin{figure}
\plotone{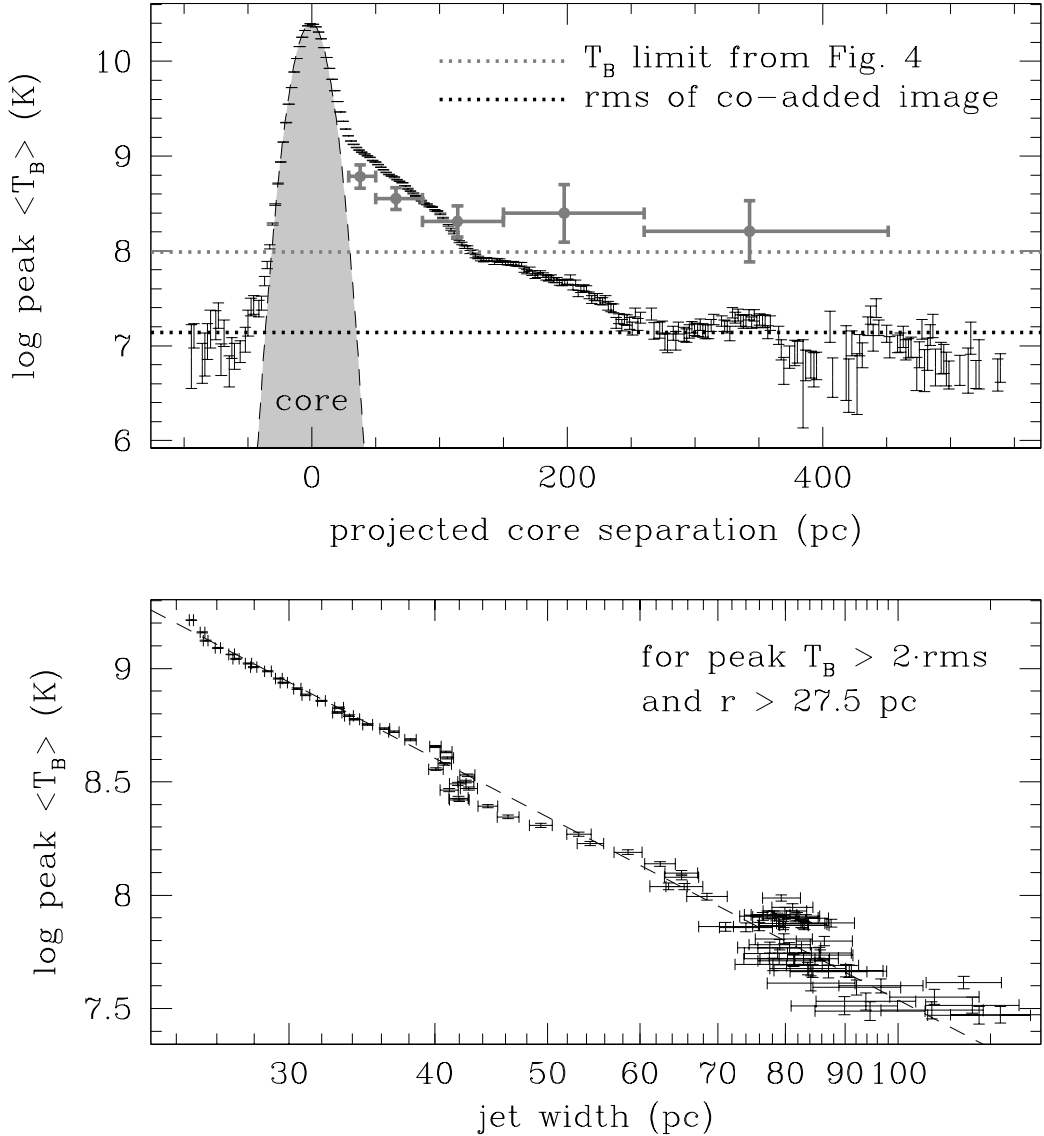}
\caption{Upper:  The peak brightness temperature from the stacked image displayed in Fig.\ \ref{coaddimg} along the jet axis as determined using Gaussian fits to the image columns.  The rms value from the image is represented by a dotted line.  The mean values for the components plotted in Fig.\ \ref{jetcomps} are plotted in grey with the adopted limiting component intensity of $10^{7.99}$ K represented by a grey dotted line.  Lower:  The brightness temperature of the jet in the stacked image versus the full width at half maximum of the jet as determined using the Gaussian fits.  These widths have been corrected for a typical dispersion of 3.1 pc about the mean jet axis that results from jet bending determined using the individual components of the sources used to make the stacked image (see \S 4.1.2).}
\label{coaddprof}
\end{figure}

To determine the peak intensity and width of the jet in the stacked image, Gaussian functions were fit to the image columns.  We have plotted the peak brightness temperatures from these Gaussian fits as a function of image position in the upper panel of Fig.\ \ref{coaddprof} along with the mean values for the components plotted in Fig.\ \ref{jetcomps}.  The plot shows the presence of a roughly Gaussian core, as expected, that extends to radii of about 27.5 pc.  Beyond this limit, the mean values from Fig.\ \ref{jetcomps} agree well with the results from the stacked image up to radii of about 150 pc.  Beyond this limit, the superior sensitivity of the stacked images allows for a detection of emission fainter than the $10^{7.99}$ K limit used for Fig.\ \ref{jetcomps} and we can see that the intensity continues to drop until the sensitivity limit of the image is reached at a radius of about 250 pc.  In the lower panel of Fig.\ \ref{coaddprof}, we have plotted the jet intensity versus the width obtained from the Gaussian fits to the image columns.  We note that most of the VIPS jets have a moderate amount of jet bending (see \S 3.2.1), which will cause the width about the mean jet axis on the stacked image to be slightly inflated.  We note that while jet components may deviate from the true jet axis to a larger degree at larger core separations, this is likely not true in practice for our analysis since we determined the orientation of each source's jet axis using a linear fit to the positions of its components on its 5 GHz image.  We have therefore estimated the magnitude of this effect by measuring the separations of the jet components from their mean jet axis for each of the sources contained in the stacked image that have more than one jet component, 18 sources in all.  For the 53 jet components of these 18 core-jet systems, the median absolute deviation of their jet axis separations is 3.1 pc.  We have therefore corrected the jet width measurements from the stacked image for an extra 3.1 pc of dispersion caused by jet bending.  The relationship between peak intensity and width for the jet on the stacked image is well approximated by a power-law with a slope of $-2.68\pm0.03$, which is quite similar to what was found for the jet of 3C 120 by \citet{wal87}.  The implications of this relationship will be discussed in \S 4.

\subsection{Jet Direction}
\subsubsection{Jet Bending on Parsec Scales}
\begin{figure}
\plotone{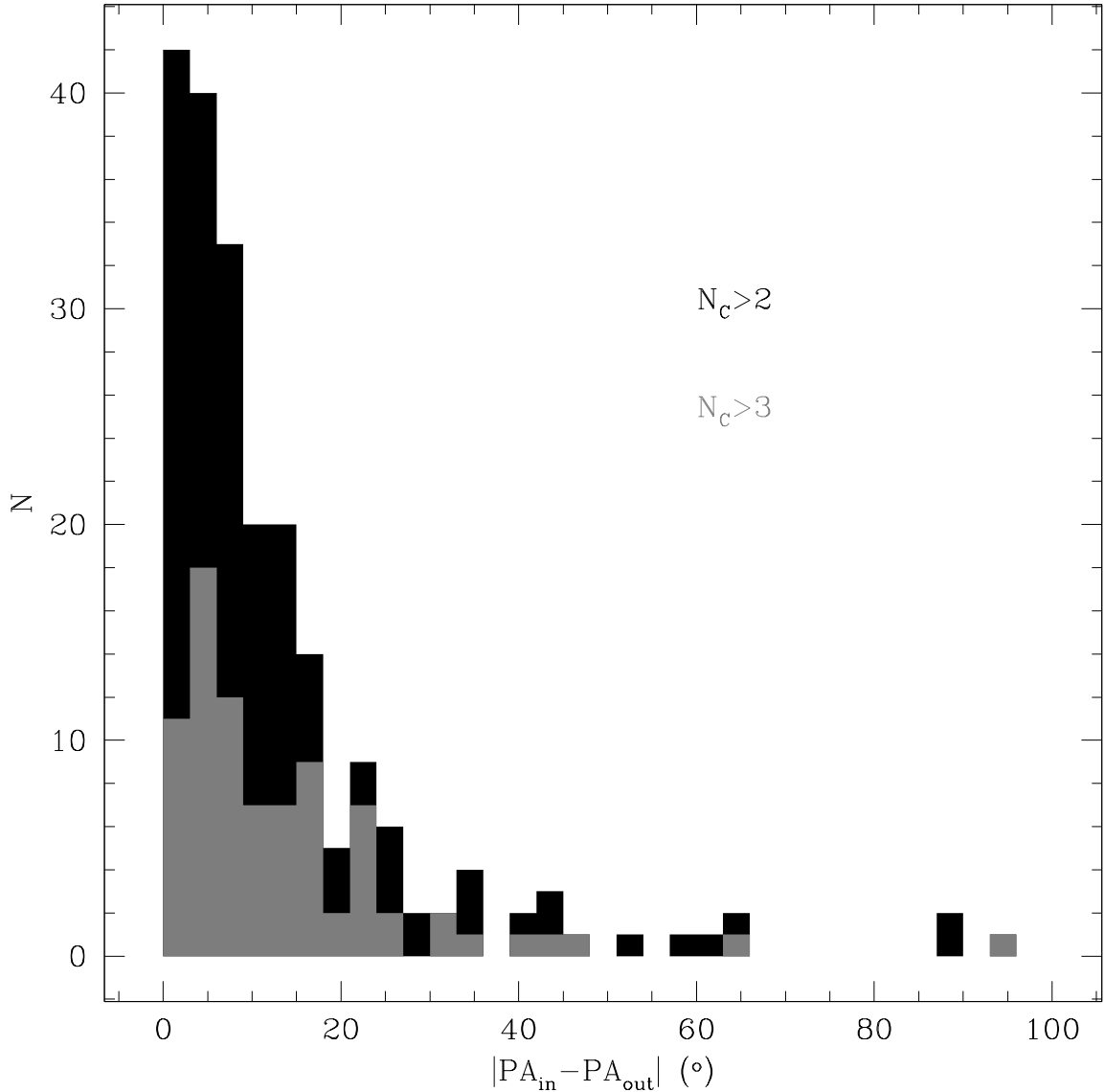}
\caption{The distribution for the absolute change in position angle from the core to the outer most jet component (see \S 3.2).  Only core-jets with at least two jet components that are separate from the core are used (i.e., $N_{C}>2$) and the distribution for core-jets with at least three jet components (i.e., $N_{C}>3$) for which the value of $|\mbox{PA}_{in}-\mbox{PA}_{out}|$ is more robustly determined (see Appendix A) is presented as a grey histogram.}
\label{angs}
\end{figure}

The projected trajectory of jets may provide important insight into how the environments of jets affect their evolution.  As a first step toward exploring this, we have derived a quantitative estimate of the amount of jet bending on parsec scales as measured using our 5 GHz images.  We have done this by developing an algorithm, detailed in Appendix A, for measuring the difference between the position angle of jet emission near the core, $PA_{in}$, and the position angle with respect to the core of the outer most jet component, $PA_{out}$, for all sources with two or more jet components, 214 sources in all.  The purpose of using an automated approach rather than a "by-eye" estimate of the amount of bending is so that a consistent, objective, and quantitative measure of the degree of jet bending could be made without being significantly affected by stray or spurious components.  The distribution for $|\mbox{PA}_{in}-\mbox{PA}_{out}|$ determined using this algorithm is displayed in Fig.\ \ref{angs}.  The distribution for sources with more than two jet components, for which the algorithm provides a better estimate of the amount of bending (see Appendix A), is also plotted separately in grey.  The peaks of the distributions appear to be between 0$^{\circ}$ and 5$^{\circ}$, implying that a modest amount of jet bending on parsec scales may be common.  However, there are instances of significant amounts of jet bending with about 20\% of sources having $|\mbox{PA}_{in}-\mbox{PA}_{out}|>20^{\circ}$ and position angle changes of up to about 90$^{\circ}$.  Using this quantitative estimate, we have found no obvious trend between the degree of jet bending and jet-to-core ratio, core or jet brightness temperature, and core or jet fractional polarization.

\subsubsection{Changes in Jet Direct from parsec to kiloparsec Scales}
\begin{figure}
\includegraphics[scale=0.9]{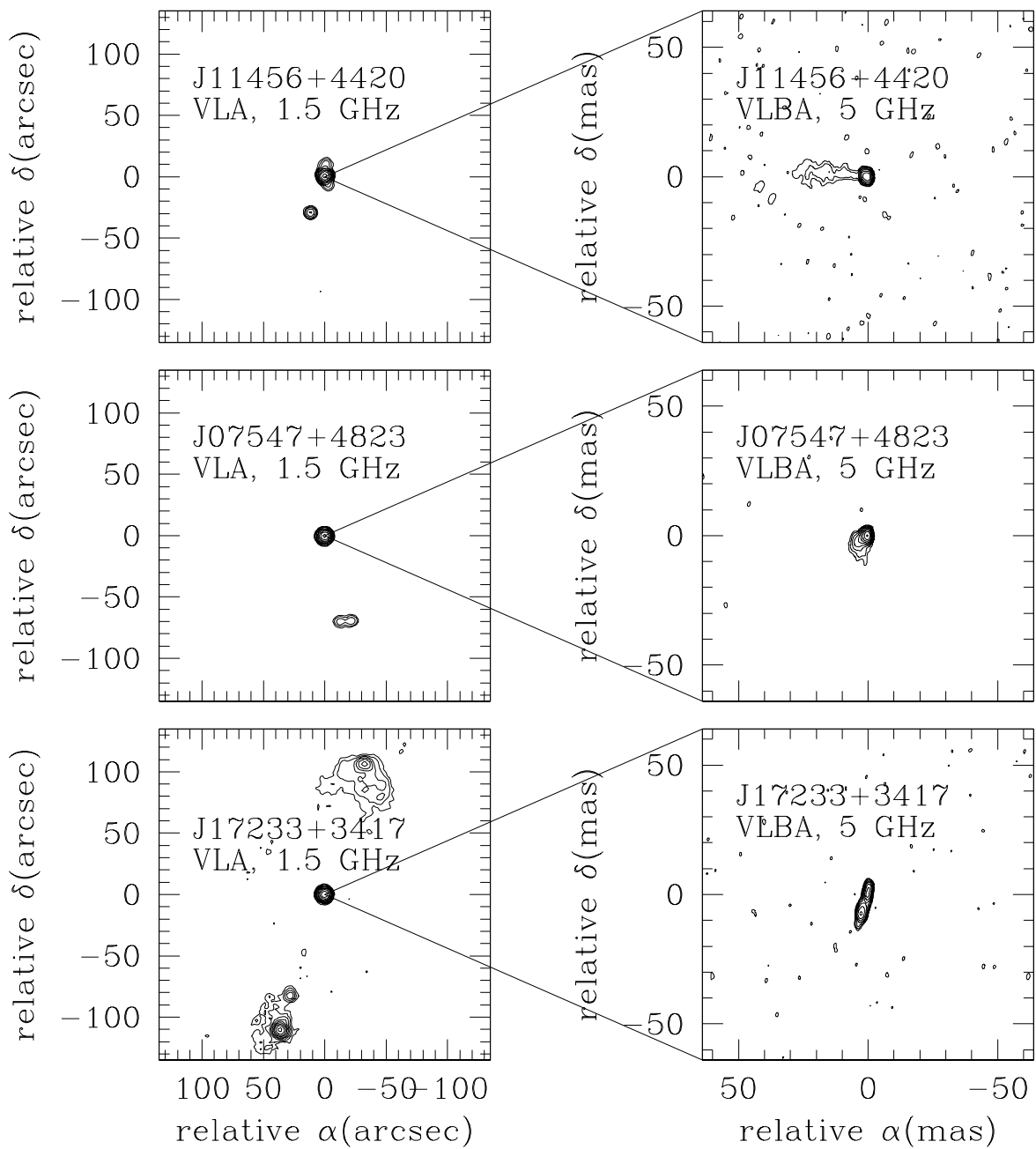}
\caption{Examples of jet and lobe emission on kiloparsec (VLA 1.5 GHz; left) and parsec (VLBA 5 GHz; right) scales.  The contour levels used in each panel are 0.9$\times$(-1, 1, 2, 4, 8, 16, 32, 64, 128, 256, 512, 1024, 2048, 4096, 8192) mJy beam$^{-1}$.}
\label{eximg}
\end{figure}

To examine how the motion of parsec-scale jet components relates to larger, kiloparsec-scale structures, we have employed VLA images of VIPS sources.  All VIPS sources have 8.5 GHz VLA images since the parent catalog for VIPS is the Cosmic Lens All Sky Survey (CLASS), an 8.5 GHz VLA imaging survey.  However, for about 98\% of VIPS sources, there are also 1.5 GHz VLA images available via the VLA FIRST survey, a relatively deep 21-cm survey of the SDSS footprint with the VLA.  Visual inspection of the CLASS and FIRST images of VIPS sources revealed that the quality of the FIRST images is typically much better (i.e., they have residual sidelobes less frequently) and that the sensitivity of the FIRST survey allows for extended emission to be present more frequently in the FIRST images than in the CLASS images.  Because of this, we have elected to use images from the FIRST survey to find and measure the properties of extended emission on kiloparsec-scales for the VIPS sources.  Three examples of VIPS sources with kiloparsec-scale structure apparent in their FIRST images are shown in Fig.\ \ref{eximg}.  These examples represent three different scenarios in which (1) the parsec-scale and kiloparsec-scale structures are oriented along completely different axes (J11456+4220), (2) the parsec-scale jet appears to be changing direction toward the kiloparsec-scale structure (J07547+4832), and (3) the parsec-scale jet and kiloparsec-scale structure are oriented along nearly the same axis (J17233+3417).\par
To quantify the degree of alignment between the parsec and kiloparsec-scale structures, we have developed an algorithm to objectively measure the orientation of extended emission on the FIRST images similar to what was used for the VIPS 5 GHz images (see Appendix B).  In the upper panel of Fig.\ \ref{vipsfirst}, we have plotted the distribution for the absolute difference between the position angles of the parsec and kiloparsec scale jet axes.  For these position angle differences, we have not taken into account whether or not the parsec or kiloparsec-scale emission is one-sided.  We have used this approach because, while all of the parsec-scale jets appear one-sided on their 5 GHz VLBA images, it is likely that this purely the result of Doppler (de-)boosting and we cannot rule out the possibility that the jets are really double-sided.  The distribution is skewed toward lower values (skewness of $\gamma_{1}=0.44$), indicating a tendency for extended emission on these two different spatial scales to be somewhat aligned.  
We note that a K-S test implies that the probability that this distribution was drawn from a flat distribution is negligible.  We also note that we see no evidence of a bimodal distribution like the one found by \citet{pea88} using similar data.  The combined VIPS and FIRST data has allowed us to measure this distribution for 132 sources, as opposed to \citet{pea88} who where restricted to using 18 sources.  It is therefore plausible that the previously observed bimodal shape was predominantly the result of small number statistics.\par
\begin{figure}
\plotone{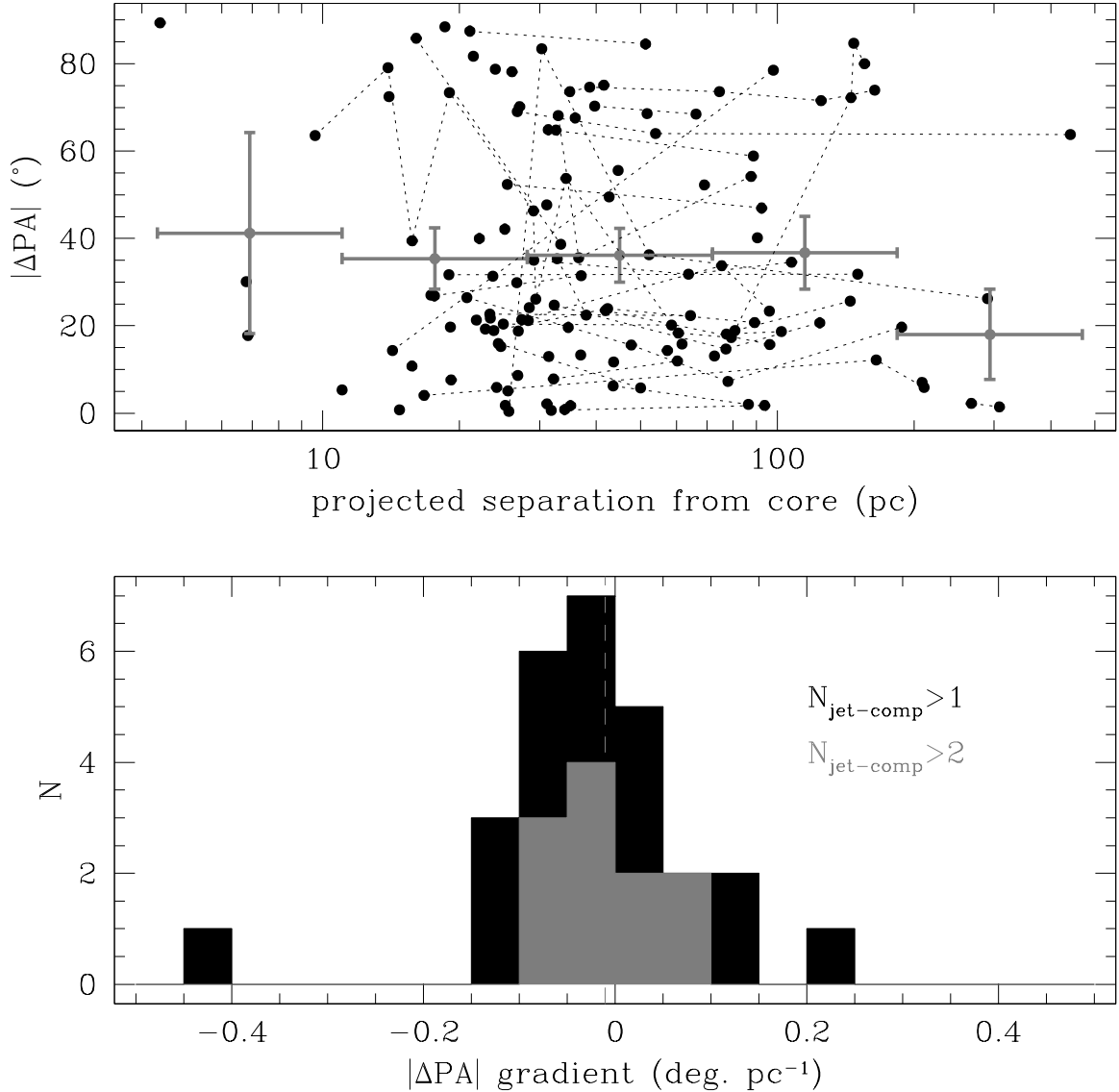}
\caption{Upper: For each individual jet component, the absolute difference between the position angle relative to the core position and the axis along which extended emission was found on larger scales from the FIRST survey images as a function of the projected separation from the core.  Core-jet systems with FIRST images that have only one component are not included.  Components belonging to the same core-jet system are connected with dashed lines.  Mean values within different bins of core separation are plotted as grey points with error bars.  Lower: The distribution of the gradients of dotted lines shown in the upper panel.  For each source with only 2 jet components, this was computed using the values for those two components.  For sources with more that 2 jet components, a line was fit to the values for the components.  Because of this, the distribution for sources with more than 2 jet components is represented separately as a grey histogram.  The medians of the two histograms are nearly identical and are represented by a vertical dashed line.}
\label{first1}
\end{figure}

The results in Fig.\ \ref{vipsfirst} imply that parsec and kiloparsec scale jets tend to be somewhat aligned, but they do not indicate the specific physical scales on which this alignment occurs.  To explore this, we have plotted absolute difference between the PAs of the jet components identified on each VIPS image and the PA of the extended emission on the corresponding FIRST image, $|\Delta\mbox{PA}|$, as a function of core separation in the upper panel of Fig.\ \ref{first1}.  We have also computed the gradient of $|\Delta\mbox{PA}|$ as a function of core separation for all sources with more than one jet component where a negative gradient implies the components are becoming more aligned with the parsec-scale emission as the move outward (i.e., $|\Delta\mbox{PA}|$ is shrinking) and a positive gradient implies the components are moving away from the axis of the parsec-scale emission.  Mean values of $|\Delta\mbox{PA}|$ within bins of core separation hint that $|\Delta\mbox{PA}|$ may be smaller for larger separations.  There is much stronger evidence for this in the distribution of $|\Delta\mbox{PA}|$ gradients displayed in the lower panel of Fig.\ \ref{first1}.  Whether all sources with more than one jet component are considered, or if only those with more than two jet components are used, the peak of the distribution is clearly $<0$.  In addition, for the 30 sources with more than one jet component, 19($\pm$4) have gradients $<0$, demonstrating that most sources have negative gradients.   This implies a tendency for parsec-scale jet components to be oriented closer to the direction of kiloparsec-scale structure at larger core separations within individual sources.  

\section{Discussion}
Using single epoch data for a large number of sources, combined with optical spectra, we have been able to statistically identify trends in the evolution of jet components.  Specifically, we have shown that the components typically expand as they move away from their cores, and that this expansion coincides with either a decrease in brightness temperature or a substantial increase in luminosity.  Relativistic hydrodynamic simulations of jets have demonstrated that the flux density of the typical jet component will remain roughly constant as it moves away from its source \citep[e.g., ][]{mid97}.  This leads us to conclude that it is likely that the potential Doppler boosting bias in Fig.\ \ref{jetcomps} is minimal and that on average, the intrinsic brightness of the components drop as their luminosites remain fairly constant.\par
This result implies that the trend between intensity and jet width seen in Fig.\ \ref{coaddprof} likely reflects a real physical relationship between these two quantities.  From this relationship, we may gain insights into how the typical jet loses or gains energy as it moves away from its core.  Following the analysis of \citet{wal87} of a similar trend observed for 3C 120, if the typical jet is assumed to be an optically thin cylinder, then the emissivity, $\epsilon$, is proportional to $T_{B} r^{-1}$, where $r$ is the jet width.  The minimum energy density of a medium with the observed emissivity, $\epsilon$, is given by $U_{min} \propto \epsilon^{4/7}$.  Our results then imply that for the typical jet, $U_{min} \propto r^{-2.10 \pm 0.02}$.  This in turn implies that the energy density per unit length has only a weak dependence on jet width, i.e., it is proportional to $r^{-0.1}$.  This is quite consistent with what was found by \citet{wal87} for 3C 120, implying that the relation (or lack thereof) between internal energy and jet width may be somewhat commonplace among AGN jets.  From this result, it appears that the typical jet loses little, if any, of its internal energy as it move away from its core, or that there is a reduction of the typical jet speed that nearly exactly balances any losses that occur.  However, as discussed above, the results plotted in Fig.\ \ref{jetcomps} imply that the mean Doppler factor does not significantly change with projected core separation for these sources, which in turn implies that the typical jet speed, $\beta$, does not change much either.  This is because \citet{ver94} and \citet{lis97} have shown that the most probable viewing angle for a jet source is $\theta \sim 0.6 \gamma^{-1}$, where the Lorentz factor $\gamma = (1-\beta^{2})^{-1/2}$.  The relation between the typical viewing angle and $\beta$ implies that in order for the mean Doppler factor to remain constant, the typical value of $\beta$ must also remain constant since $\delta$ is only a function of $\theta$ and $\beta$.  Consequently, we conclude that it is more likely that for these sources, little energy per unit length is lost as their jets move outward.  Combining this results with the fact that the integrated luminosity density of the typical jet component remains roughly constant (see \S 3.1.1 and Fig.\ \ref{jetcomps}) as it expands, implies that the bulk of this expansion is likely perpendicular to the jet.\par
In addition to this result, we have also demonstrated that the jet components on parsec scales tend be aligned with the direction of extended emission on kiloparsec scales.  We have also found that this alignment improves as the jet components move further from their cores on projected scales as small as $\sim$50-100 pc.  These results suggest that interactions with the surrounding medium may produce realignment of the jets on relatively small projected scales.  We note that our data do not provide the opportunity to thoroughly study the nature of this realignment to discern whether it consists of relatively gradual changes in jet direction, as with 3C 120 \citep{wal87}, or of more abrupt changes similar to the event recently observed for 3C 279 \citep{hom03}.  However, the fact that our sample shows evidence of relatively modest amounts of jet bending on projected scales of $\sim$100 pc suggests that the former scenario is more common than the latter.

\acknowledgements  The authors would like to thank the referee for useful comments and suggestions.  This research was performed while the lead author held a National Research Council Research Associateship Award at the Naval Research Laboratory.  Basic research in astronomy at the Naval Research Laboratory is supported by 6.1 base funding.  The National Radio Astronomy Observatory is a facility of the National Science Foundation operated under cooperative agreement by Associated Universities, Inc.

\appendix
\section{Jet-bending}
\begin{figure}
\plotone{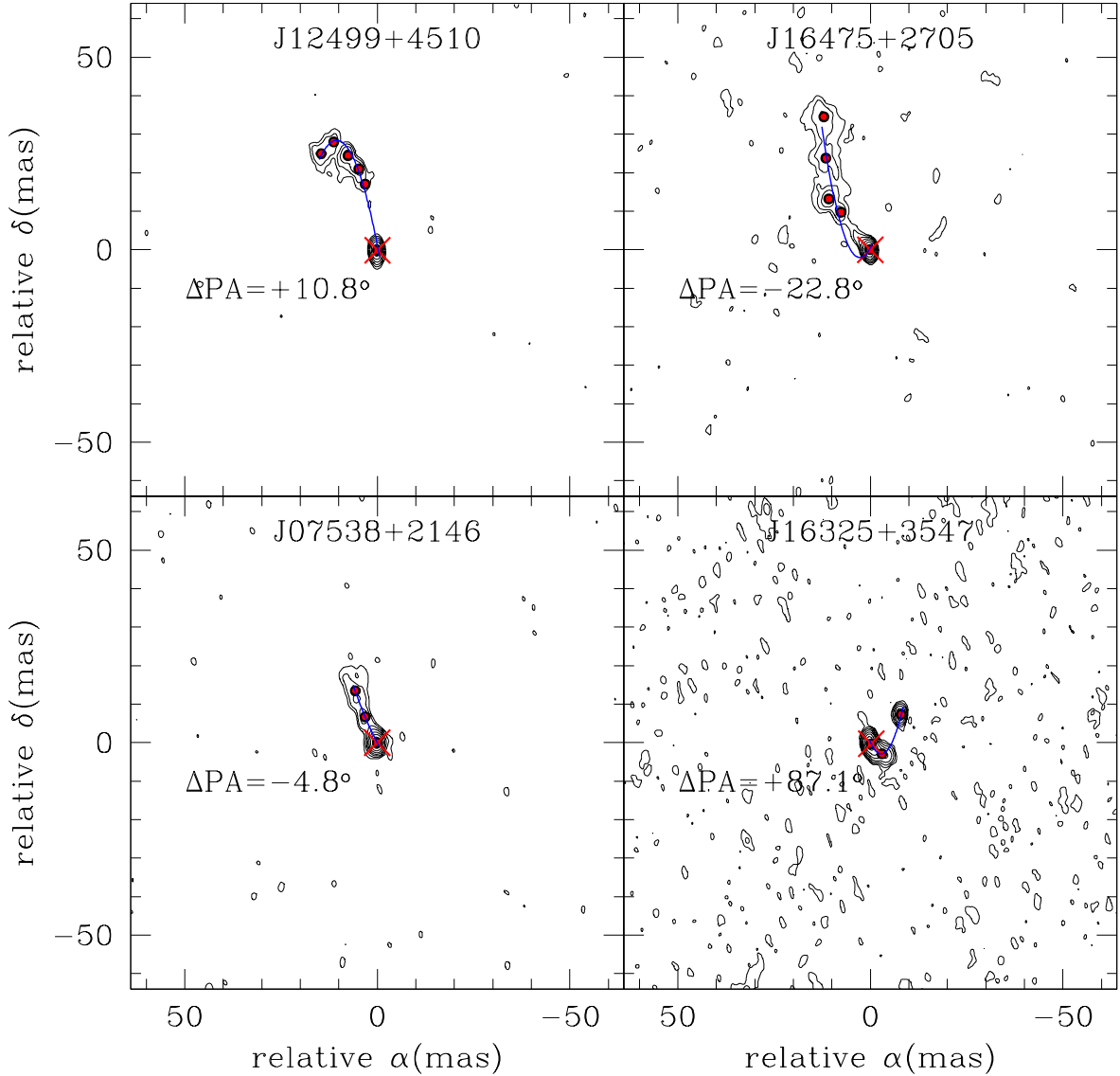}
\caption{Examples of the parabola-fitting procedure used to quantify the amount of jet bending described in \S 3.2.  In all panels, the contours were made using the total intensity 5 GHz VLBA images with the same levels as were used in Fig. \ref{eximg}.  Each core center is marked with a red $\times$ symbol and the jet component centers are marked with red points; the parabolic fits are plotted as blue curves.  The top panels show examples of two sources with more than three components; examples of sources with only three components are show in the bottom panels.  For the sources in the left panels, the inner position angles were measured using the slopes of the parabolic fits at the core positions.  In each of the panels on the right, the extreme of the parabolic fit lies in between the core and the nearest jet component, and the inner position angle was computed using the closest component (see \S Appendix A).}
\label{jbend}
\end{figure}

To establish an objective and consistent quantitative measure of the amount of jet bending on parsec scales, we have developed the following algorithm which uses our 5 GHz images.  First, for each core-jet with more than two jet components, a parabola was fit to the positions of the core and the jet components in order to asses how much the jet bends.  For sources with only two jet components, this was formally not a fit, but rather an exact solution of the parabola that passes through the center of the core and each jet component.  There are 214 VIPS core-jet systems with two or more jet components for which this could be done, 104 of which have redshifts.  For each of these core-jets, the slope of the parabola at the position of the core was used to determine the inner position angle of the jet, PA$_{in}$.  The value of the parabola at the right ascension of the jet component farthest from the core was then used to compute the outer jet position angle relative to the position of the core, PA$_{out}$.  It was found that in about 15\% of cases, the extreme of the fitted parabola was in between the position of the core and the closest jet component.  In these cases, the difference between PA$_{in}$ and PA$_{out}$ is artificially large since the fitted parabola makes a significant turn in direction in a region devoid of any components.  Since this is not the case for a large fraction of the sources, we have elected not to abandon our parabola fitting method because of the motivations stated above.  Instead, we have chosen to take PA$_{in}$ to be the position angle of the nearest jet component relative to the core position in these few (32 out of 214) instances.  Examples of the parabolic fits can be seen in Fig.\ \ref{jbend}.

\section{Kiloparsec-scale Position Angles}
To quantitatively asses the degree of alignment between the parsec-scale extended emission and the kiloparsec-scale emission, we have used the following algorithm to measure the PA of the extended emission on the FIRST images.  For consistency, this algorithm was based on the routine used to measure the position angle on parsec-scales from the VIPS 5 GHz images \citep[see ][]{hel07}.  Using the AIPS task SAD, Gaussian components were fit to significant peaks in each FIRST image down to the 6$\sigma$ level.  As with the VIPS sources, the dominant Gaussian components within each FIRST image were identified as the brightest components whose combined flux density is $>$95\% of the flux density of all Gaussian components.  For each image with more than one component, a line was fit to the component positions and was used to measure a kiloparsec-scale position angle.  Each image with more than one component was then visually inspected to flag those sources whose morphology appeared roughly circularly symmetric and/or not oriented along a singular axis.  About 7\% of the sources with multiple components were flagged by this inspection.  For the remaining 93\% that were not flagged, we measured the absolute difference between the position angle of each parsec-scale jet component measured with respect to its core and the axis along which kiloparsec-scale structure was found in the FIRST image, referred to as $|\Delta\mbox{PA}|$.  

{}

\clearpage

\clearpage

\clearpage


\clearpage

\clearpage


\begin{figure}
\plotone{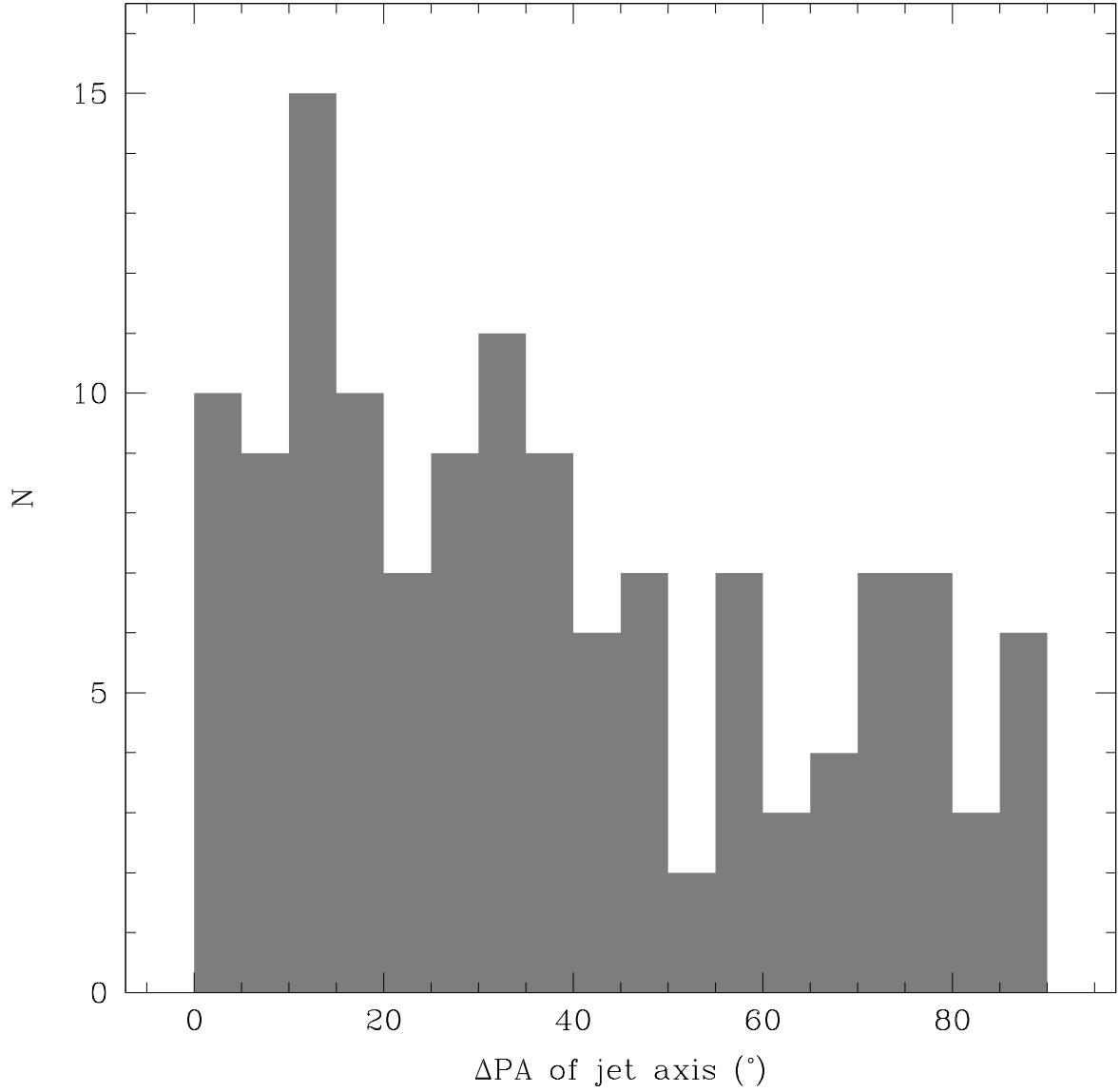}
\caption{The distribution of the absolute difference between the position angle of the parsec-scale jet emission measured from the VIPS images to that of the kiloparsec-scale emission measured from the FIRST images.}
\label{vipsfirst}
\end{figure}

\clearpage

\clearpage

\end{document}